\documentclass[iap,preprint,showpacs,superscriptaddress,groupedaddress,balancelastpage,floatfix]{revtex4-2}  
\usepackage[utf8]{inputenc}
\usepackage{graphicx,dcolumn,bm,amssymb,amsmath,amsfonts,xcolor}
\usepackage{booktabs}
\usepackage[
pdfstartview=FitBV,
bookmarks=true,
bookmarksopen=true,
colorlinks =true,
linkbordercolor=blue,
linkcolor = blue,
citecolor = blue,
urlcolor=blue]{hyperref}
\usepackage[capitalize]{cleveref} 

\crefname{section}{Sec.}{Sections}
\crefname{table}{TABLE}{TABLEs.}
\crefname{figure}{FIG.}{FIGs.}

\begin{document}

	\title{Hyperradial distribution function of few-body problems: a new arena for extreme value theory}
	
	\author{Yu Wang}
	\email{yu.wang.5@stonybrook.edu}
	\affiliation{Department of Physics and Astronomy, Stony Brook University, Stony Brook, NY 11794, USA}
	\author{Marjan Mirahmadi}
	\email{mirahmadi@fhi-berlin.mpg.de}
	\affiliation{Fritz-Haber-Institut der Max-Planck-Gesellschaft, Faradayweg 4-6, D-14195 Berlin, Germany}
	\author{Ahmed A. Elkamshishy}
	\email{aelkamsh@purdue.edu}
	\affiliation{ Department of Physics and Astronomy, Purdue University, West Lafayette, IN 47907, USA}
	\author{Jes\'{u}s P\'{e}rez-R\'{i}os}
	\email{jesus.perezrios@stonybrook.edu}
	\affiliation{Department of Physics and Astronomy, Stony Brook University, Stony Brook, NY 11794, USA}
	\affiliation{Institute for Advanced Computational Science, Stony Brook University, Stony Brook, NY 11794, USA}
	\affiliation{Fritz-Haber-Institut der Max-Planck-Gesellschaft, Faradayweg 4-6, D-14195 Berlin, Germany}

	\begin{abstract}
		This work explores classical capture models for few-body systems via a Monte Carlo method in hyperspherical coordinates. In particular, we focus on van der Waals and charged-induced dipole interactions. As a result, we notice that, independently of the number of particles and interparticle interaction, the capture hyperradial distribution function follows a Fréchet distribution, a special type of the generalized extreme value distribution. Besides, we elaborate on the fundamentals of such universal feature using the general extreme value theory, thus, establishing a connection between extreme value theory and few-body physics.  
	\end{abstract}
	\maketitle
	
	\section{Introduction}\label{sec1}
	Classically, when the interaction potential energy is comparable to the colliding partners' kinetic energy, the trajectory drawn by the particles deviates from the uniform rectilinear motion, leading to a collision~\cite{Levine2005,Perez-Rios2020}. The distance at which this occurs defines the range of the interaction, also known as the capture radius in the language of classical capture models. Then, using the capture radius is possible to define scattering observables such as cross section and reaction rate~\footnote{This is strictly true in the case of zero impact parameter, or null angular momentum collision.}. A prime example of a classical capture is the well-known Langevin rate for charged-neutral collisions~\cite{Langevin1905}, although it includes the effects of the centrifugal barrier. 
	
	

	Recently, classical capture models have been introduced within the framework of direct three-body recombination processes A + A + A $\rightarrow$ A$_2$ + A~\cite{Perez-Rios2014,Perez-Rios2018,Perez-Rios2021,Mirahmadi2021,Mirahmadi2021b}, also known as ternary association reactions, i.e., the formation of a molecule from a collision of three free atoms. In this context, direct means that the reaction occurs without invoking the existence of an intermediate two-body complex that the third body may stabilize or dissociate. In particular, after using hyperspherical coordinates to characterize the configuration of the particles, in which the hyperradius emerges as a natural generalization of the radius in spherical coordinates, a distribution of capture hyperradius is found for given collision energy. Moreover, and surprisingly enough, the distribution of capture hyperradius follows a generalized extreme value (GEV) distribution~\cite{Singh1998} independently of the collision energy. However, it has not yet explained why a GEV distribution appears. Also, classical capture models have not been introduced in the framework of four-body collisions and beyond.
	

	In this work, we compute the capture hyperradius distribution for three- four- and five-body collisions using different interatomic interactions through a Monte Carlo approach sampling the different initial particle configurations of the N-body system. As a result, we find the GEV distribution in all studied scenarios independently of the collision energy, the number of particles, and interparticle interaction. Therefore, our results indicate that the shape of the hyperradial distribution function is a universal property of few-body systems. Additionally, we find that the extreme value theory applies to few-body systems and is key to understanding hyperradial distributions. The paper is organized as follows: \cref{sec2} presents our methodology to extract the hyperradial distribution function; \cref{sec3} shows our results that are further analyzed and discussed in \cref{sec4}. Finally, a summary of our chief findings and conclusions are discussed in \cref{sec5}.
	
	\section{Theoretical framework}\label{sec2}
	
	The classical capture radius is estimated as the distance for which the interaction potential energy is of the same order as the kinetic energy of a scattering system. Therefore, it is possible to find the classical capture radius for N-body processes.
	
	\subsection{Interaction potential in N-body systems}
	Let us assume a system of N particles with masses $m_i(i=1,...,N)$ located at positions $\vec{r}_i(i=1,...,N)$ interacting through a pair-wise potential $U(r_{ij})$, wherein the relative distance $r_{ij}$ is 
	\begin{equation}
		r_{ij}=\|\vec{ r}_j-\vec{ r}_i\|,
	\end{equation}
	such that the total interaction potential of the system reads as
	\begin{equation}
		\label{eq2}
		U(\vec{r}_1,...,\vec{ r}_N)=\sum_{i=1}^{N}\sum_{j> i}^{N}U(r_{ij}).
	\end{equation}
	Next, as customary on treating few-body systems from a chemical physics standpoint, we introduce the Jacobi coordinates given as	
	\begin{eqnarray}
		\label{eq3}
		\vec{\rho}_1&=&\vec{r}_2-\vec{r}_1\nonumber \\
		\vec{\rho}_2&=&\vec{r}_3-\frac{m_1\vec{r}_1+m_2\vec{r}_2}{m_1+m_2} \nonumber \\
		&&\vdots\\
		\vec{\rho}_{N-1}&=&\vec{r}_N-\frac{\sum_{i=1}^{N-1}m_i\vec{r}_i}{\sum_{i=1}^{N-1}m_i} \nonumber\\
		\vec{\rho}_{\text{CM}}&=&\frac{\sum_{i=1}^{N}m_i\vec{r}_i}{\sum_{i=1}^{N}m_i}\nonumber
	\end{eqnarray}
	which lead to the interaction potential $U(\vec{ \rho}_1,...,\vec{ \rho}_{N-1})$. Therefore, to determine the characteristic interaction range at a given collision energy $E_{\text{col}}$, we need to solve the equation $U(\vec{ \rho}_1,...,\vec{ \rho}_{N-1})=E_{\text{col}}$. By doing so, we find different particle configurations associated with  the same collision energy. This leads to a complecated distribution of interparticle distances which is hard to analyze. 
	
\begin{figure}[h]
		\centering
		\includegraphics[scale=0.58]{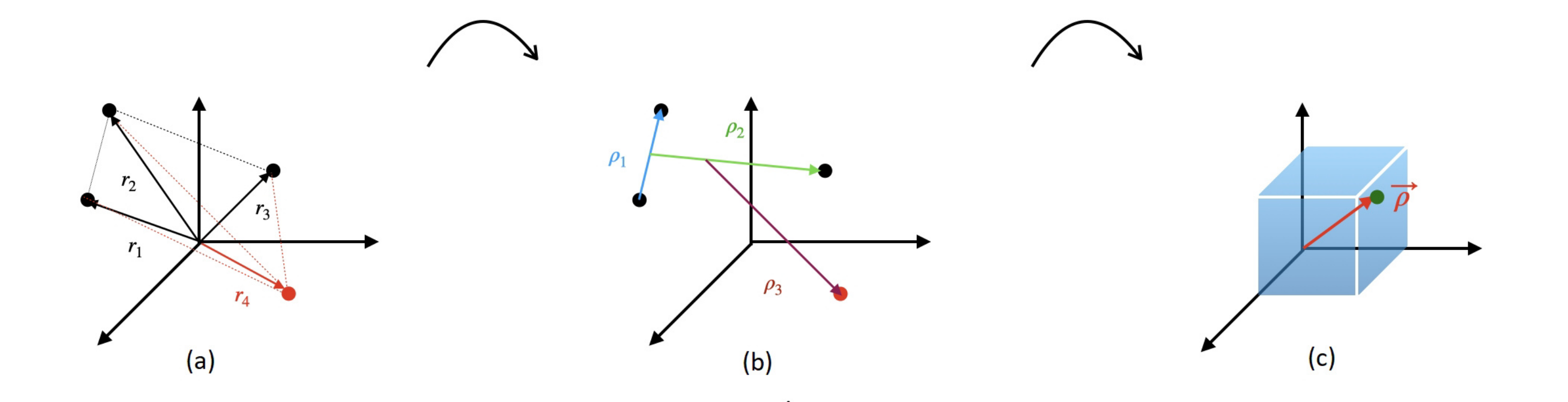}
		\caption{Four-body interactions. Panel (a) shows the position of four particles in a 3-dimensional space, panel (b) describes the same particles but in Jacobi coordinates, and panel (c) shows a single particle in a 9-dimensional space.}
		\label{fig1}
	\end{figure}

	It is well-known that the dynamics of an N-body problem in a 3-dimensional space can be mapped onto a single particle dynamics in a D-dimensional space, with $D=3\times N-3$, i.e., three spatial degrees of freedom per particle minus the degrees of freedom associated with the trivial center of mass motion. In particular, following the pioneering work of Smith and Shui~\cite{Smith1960,Smith1962,Shui1972,Shui1973}, it is possible to represent the position of N particles of the system via a single D-dimensional vector given by
	
	\begin{equation}
		\label{eq4}
		\vec{\rho}= \begin{pmatrix}
			\vec{\rho}_1\\
			\vec{\rho}_2\\
			\vdots \\
			\vec{\rho}_{N-1}
		\end{pmatrix},
	\end{equation}
	as illustrated for a four-body system in \cref{fig1}. As a result, the interaction potential of the system is given by $U(\vec{\rho})$.
	
	\subsection{Hyperspherical coordinates}
	Hypershperical coordinates are a natural generalization of spherical coordinates into higher dimensional space. In particular, following Avery's definition of hyperangles~\cite{Avery2012}, the components of a D-dimensional vector are expressed as
	
	\begin{eqnarray}
		\label{eq5}
		&&x_{1}=\rho \sin{\alpha_1}\sin{\alpha_2}...\sin{\alpha_{D-3}}\sin{\alpha_{D-2}}\sin{\alpha_{D-1}}\nonumber\\
		&&x_{2}=\rho \cos{\alpha_1}\sin{\alpha_2}...\sin{\alpha_{D-3}}\sin{\alpha_{D-2}}\sin{\alpha_{D-1}}\nonumber\\
		&&x_{3}=\rho \cos{\alpha_2}...\sin{\alpha_{D-3}}\sin{\alpha_{D-2}}\sin{\alpha_{D-1}}\nonumber\\
		&&\vdots\\
		&&x_{D}=\rho\cos{\alpha_{D-1}}\nonumber
	\end{eqnarray}
	where $0\le\alpha_1< 2\pi$ and $0\le \alpha_{i\ne1}\le \pi$ stand for the D-1 hyperangles, whereas $\rho$ is the hyperradius. Similarly, the infinitesimal volume element is given by
	
	\begin{equation}
		\label{eq6}
		d\tau=\rho^{D-1} d\rho \prod^{D-1}_{i=1}\sin^{i-1}\alpha_i d\alpha_i.
	\end{equation}
	
	\subsection{Capture hyperradial distribution for N-body systems}
	
	Via \cref{eq3}, it is possible to express the position of the particles ($\vec{r_i}$) as a function of the Jacobi vectors, and with it, the interparticle distance. Next, using hyperspherical coordinates to describe the D-dimensional space (see \cref{eq5}), we find 
	
	
	\begin{equation}
		U(\vec{\rho})=U(\rho,\vec{\alpha})=\sum_{i=1}^{N}\sum_{j> i}^{N}U(r_{ij}(\rho,\vec{\alpha}))
	\end{equation}
	where $\vec{\alpha}=(\alpha_1,...,\alpha_{D-1})$. Finally, solving 
	
	\begin{equation}
		\label{eq8}
		U(\rho,\vec{\alpha})=E_{\text{col}}, 
	\end{equation}
	leads to the capture hyperradius for a given set of hyperangles. Indeed, every configuration controlled by the hyperangles gives a different hyperradius, leading to a capture hyperradial distribution rather than a single capture hyperradius.

	\section{Methods}\label{sec3}
	Herein, we explore N-body collisions in which one particle (A) is different from the rest (B). In particular, we envision a system of N particles interacting via van der Waals (vdW) interactions described by 
	
	\begin{equation}
		\label{eq9}
		U(\rho,\vec{\alpha})=-\sum^{N-1}_{i\ne n}\frac{C^{A-B}_6}{r^{6}_{ni}(\rho,\vec{\alpha})}-\sum^{N-1}_{i\ne n}\sum^{N-2}_{j\ne n>i} \frac{C^{B-B}_6}{r^{6}_{ij}(\rho,\vec{\alpha})},
	\end{equation}
	where $C_6^{A-B}$ and $C_6^{B-B}$ stand for the van der Waals coefficient between A and B particles and B-B particles, respectively. In \cref{eq9} it is assumed that the A particle is placed at $\vec{r}_n$. Indeed, the choice for labeling the particles in the system does not affect the results since we are interested in the energy landscape.

	The second case under study is based in ion-neutral interactions, in which the A particle is positive charged whereas B particles are neutral. Then, the interaction potential reads as
	\begin{equation}
		\label{eq10}
		U(\rho,\vec{\alpha})=-\sum^{N-1}_{i\ne n}\frac{C^{A-B}_4}{r^{4}_{ni}(\rho,\vec{\alpha})}-\sum^{N-1}_{i \ne n}\sum^{N-2}_{j\ne n>i} \frac{C^{B-B}_6}{r^{6}_{ij}(\rho,\vec{\alpha})},
	\end{equation}
	wherein $C_4^{A-B}$ is the charged induced dipole long-range interaction coefficient, that only depends on the polarizaiblity of the atoms. 
	
	We use a Monte Carlo approach to generate the capture hyperradial distribution by generating random hyperangles. In each Monte Carlo step, we generate a set of hyperangles via the inverse transformation method to generate random numbers based on its probability density function~\cite{landau_binder_2009,MC2}. This, is obtained from the hyperangular part of the differential volume element given by \cref{eq6} normalized via a constant $k_i$ associated with $\alpha_i$ as
	
	
	\begin{equation}
		\label{eq11}
		k_i=\left\{
		\begin{array}{cr}
			\dfrac{1}{2\pi}, &  \quad i = 1 \\
			& \\
			\dfrac{1}{\int^{\pi}_{0}\sin^{i-1}\alpha_i d\alpha},  &  \quad i > 1 \\
			\end{array} \right.
	\end{equation}

	Next, in virtue of \cref{eq4,eq5} and using the generated $\vec{\alpha}$, we find the Jacobi vectors associated with the generated configuration. Then, solving \cref{eq3} for $\vec{r}_i$ (i=1,$\dots$,N) allows us to obtain the interparticle vectors $\vec{r}_{ij}(\rho,\vec{\alpha})$. With this, we have all the ingredients to solve $U(\rho,\vec{\alpha})=E_{\text{col}}$ for $\rho$. In this way, for each Monte Carlo step we obtain a capture hypperradius associated with $E_{\text{col}}$.
	Note that the interaction coefficients in \cref{eq9,eq10} are considered as constant parameters in the calculations. 
	
	\section{Results and Discussion}\label{sec4}
\begin{figure}[t]
		\centering
		\includegraphics[scale=0.35]{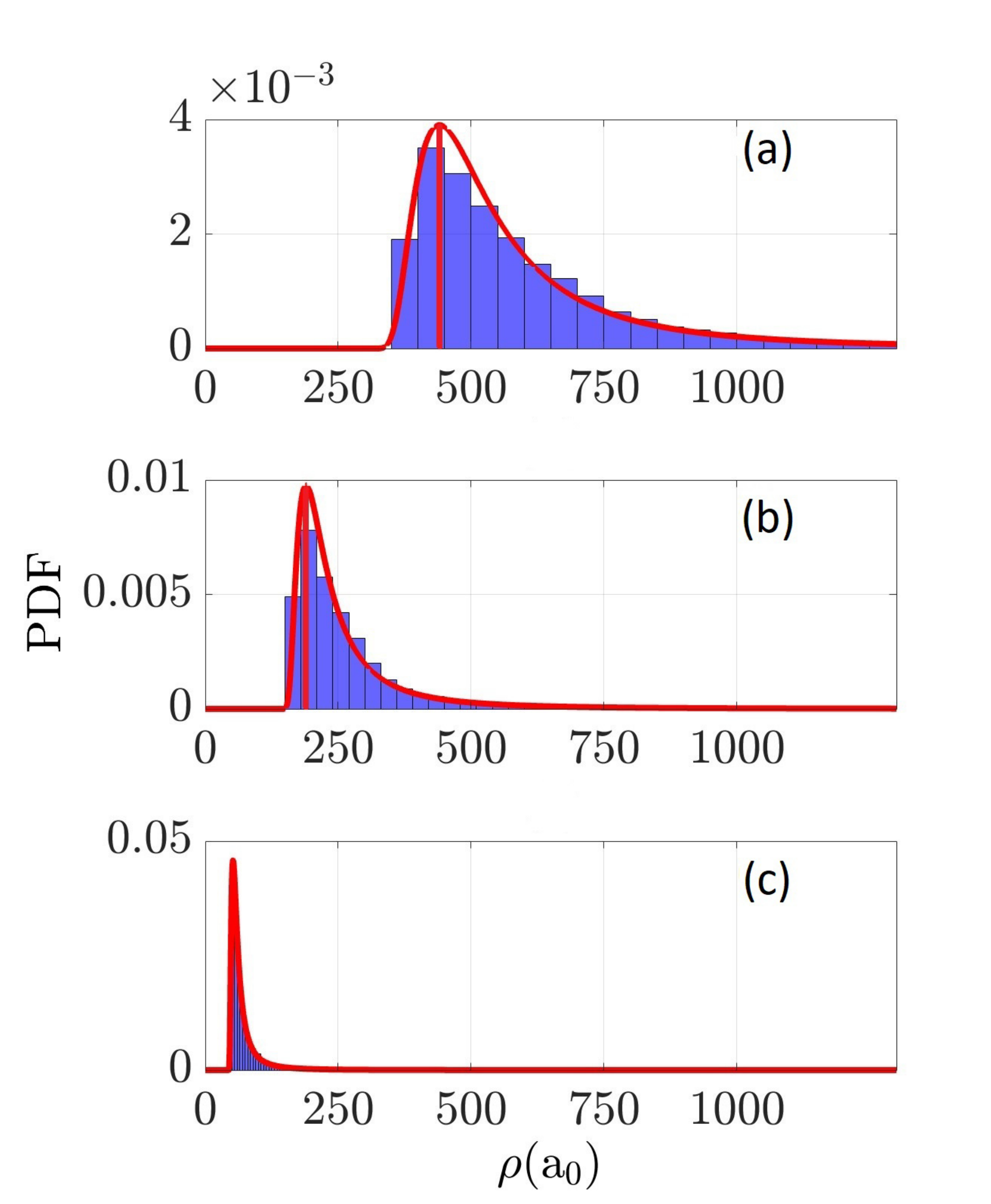}
		\caption{Hyperradius distribution for a collision energy of 1~mK. Probability density function (PDF) associated with N$^{+}$-Ar-Ar [panel (a)], N$^{+}$-He-He [panel (b)] and N-He-He [panel (c)]. The solid lines represent the fitting to a GEV distribution, whereas the mode is indicated by the vertical lines: are 440~a$_0$, 203~a$_0$ and 52~a$_0$ for panels (a), (b) and (c), respectively.  }
		\label{fig2}
	\end{figure}

	The capture hyperradial distribution for N$^+$-Ar-Ar, N$^+$-He-He and N-He-He three-body collisions at $E_\text{col}=1$~mK have been obtained and the results are displayed in \cref{fig2}. Here, we use $C_4^\text{{N-He}}=0.69$~a.u., $C_4^{\text{N-Ar}}=5.54$~a.u., $C_6^{\text{He-He}}=1.35$~a.u.~\cite{C6He2}, $C_6^{\text{Ar-Ar}}=64.3$~a.u.~\cite{C6Ar2}, and $C_6^{\text{N-He}}=5.7$~a.u.~\cite{C6NHe}, to calculate the capture hyperradius. In \cref{fig2}, it is noticed that three-body processes involving charged-neutral interactions lead to a broader hyperradial distribution than in the case of solely vdW interactions. Similarly, the mode of the distributions involving a charged particle occurs at a larger hyperradius than in the case of vdW interactions. As expected, the mode and the width of the distribution are attributed to the stronger long-range nature of charge-neutral interactions ($1/r^4)$ versus vdW forces ($1/r^6)$. Furthermore, it turns out that the hyperradial distribution  is always described by a GEV distribution with the probability density function (PDF) given by (depicted by the red solid line in the figures)
	
	\begin{equation}
		\label{eq12}
		f(\rho) = \frac{1}{\delta}\exp\left[-\left(1+\xi\frac{\rho-\beta}{\delta}\right)^{-\frac{1}{\xi}}\right]\left(1+\xi\frac{\rho-\beta}{\delta}\right)^{-1-\frac{1}{\xi}}, 
	\end{equation}
	with $1+\xi\frac{\rho-\beta}{\delta}>0$. Here, $\xi$ is the shape parameter that is directly related to the assignment of different types of GEV distribution: Weibul, Fréchet and Gumbel, for $\xi=0$, $\xi>0$ and $\xi<0$, respectively; $\delta$ is the scale parameter and $\beta$ indicates the location parameter. An extensive listing of these parameters and their values for different systems is shown in \cref{table1}.

\begin{figure}[t]
		\centering
		\includegraphics[scale=0.35]{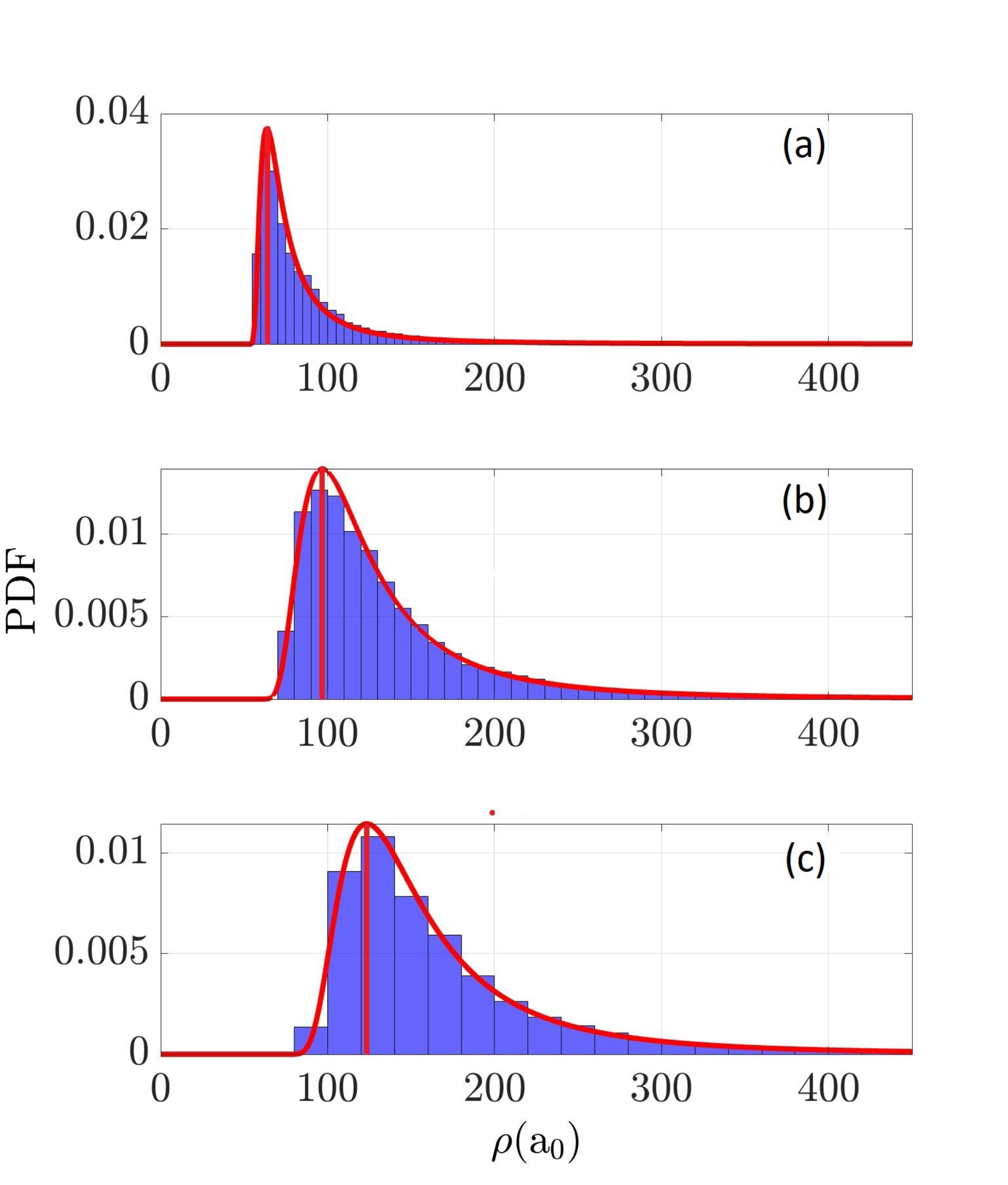}
		\caption{Hyperradius distribution for a collision energy of 1~mK as a function of the number of colliding particles. Panel (a) shows the Probability density function (PDF) for As-He-He, whereas panel (b) and (c) show the results for As-He-He-He and As-He-He-He-He, respectively. The solid lines represent the fitting to a GEV distribution, whereas the mode is indicated by the vertical lines: 64~a$_0$, 92~a$_0$ and 123~a$_0$ for panels (a), (b) and (c), respectively.}
		\label{fig3}
	\end{figure}

	To study the role of the number of particles on the hypperradial distribution, we have computed the hyperradial distribution for As-He-He, As-He-He-He and As-He-He-He-He. In particular, we include vdW-type interactions following \cref{eq9}, wherein A=As and B=He. Employing $C_6^{\text{As-He}}=17.16$~a.u~\cite{C6NHe} and $C_6^{\text{He-He}}=1.35$~a.u.~\cite{C6He2}, we obtain the results shown in \cref{fig3}. Analogously to three-body collisions shown in \cref{fig2}, the hyperradial distributions are well described via a GEV distribution. However, in this case, the width of the distribution shows a clear dependence on the number of particles whereas the mode shows a more subtle dependence. By comparing the changes in  panels (b) and (c) of \cref{fig2} and \cref{fig3}, we notice that, even though the mode and width of the hyperradial distribution depend on both number of particles and interparticle interactions, they are more sensitive to the changes in the latter one.

\begin{figure}[h]
		\centering
		\includegraphics[scale=0.35]{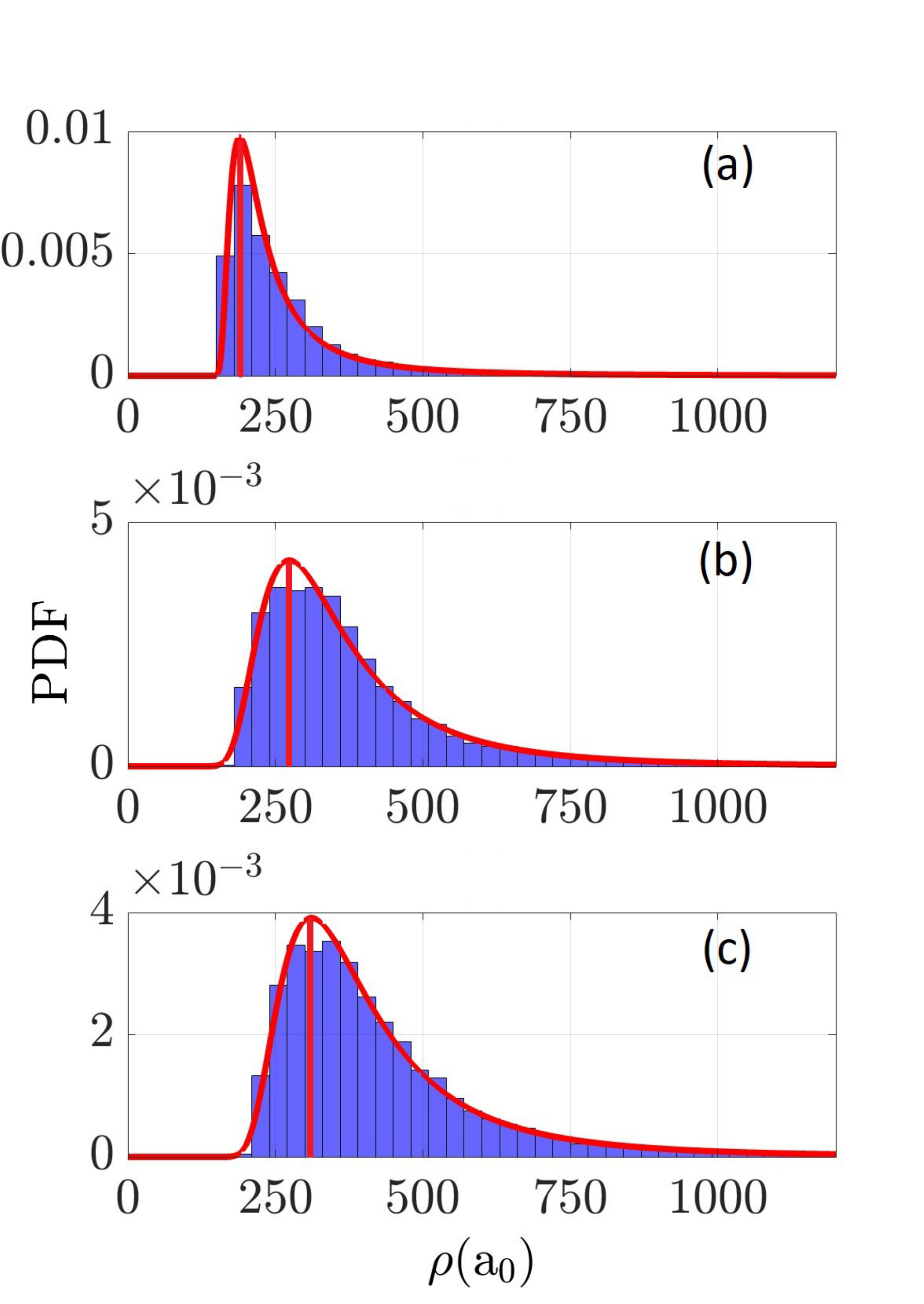}
		\caption{Hyperradius distribution for a collision energy of 1~mK as a function of the number of colliding particles. Panel (a) shows the Probability density function (PDF) for N$^+$-He-He, whereas panel (b) and (c) show the results for N$^+$-He-He-He and N$^+$-He-He-He-He, respectively. The solid lines represent the fitting to a GEV distribution, whereas the mode is indicated by the vertical lines: 201~a$_0$, 273~a$_0$ and 311~a$_0$ for panels (a), (b) and (c), respectively.}
		\label{fig4}
	\end{figure}

	Similarly, \cref{fig4} shows our results for the hyperradial distribution for three- four- and five-body collisions involving charged-neutral interactions following \cref{eq10}, wherein A=N$^+$ and B=He with $C_4^{\text{N-He}}=0.69$~a.u and $C_6^{\text{He-He}}=1.35$~a.u.~\cite{C6He2}. As in previous cases, a GEV describes the computed distribution, in which the width and the mode depend on the number of colliding particles, although, here the relationship is weak in comparison with vdW-type interactions. In other words, the number of particles has a minimum effect on the hyperradial distribution when one of them is a charged particle in comparison with the scenario of vdW-type interactions. 
	

\begin{figure}[h]
		\centering
		\includegraphics[scale=0.4]{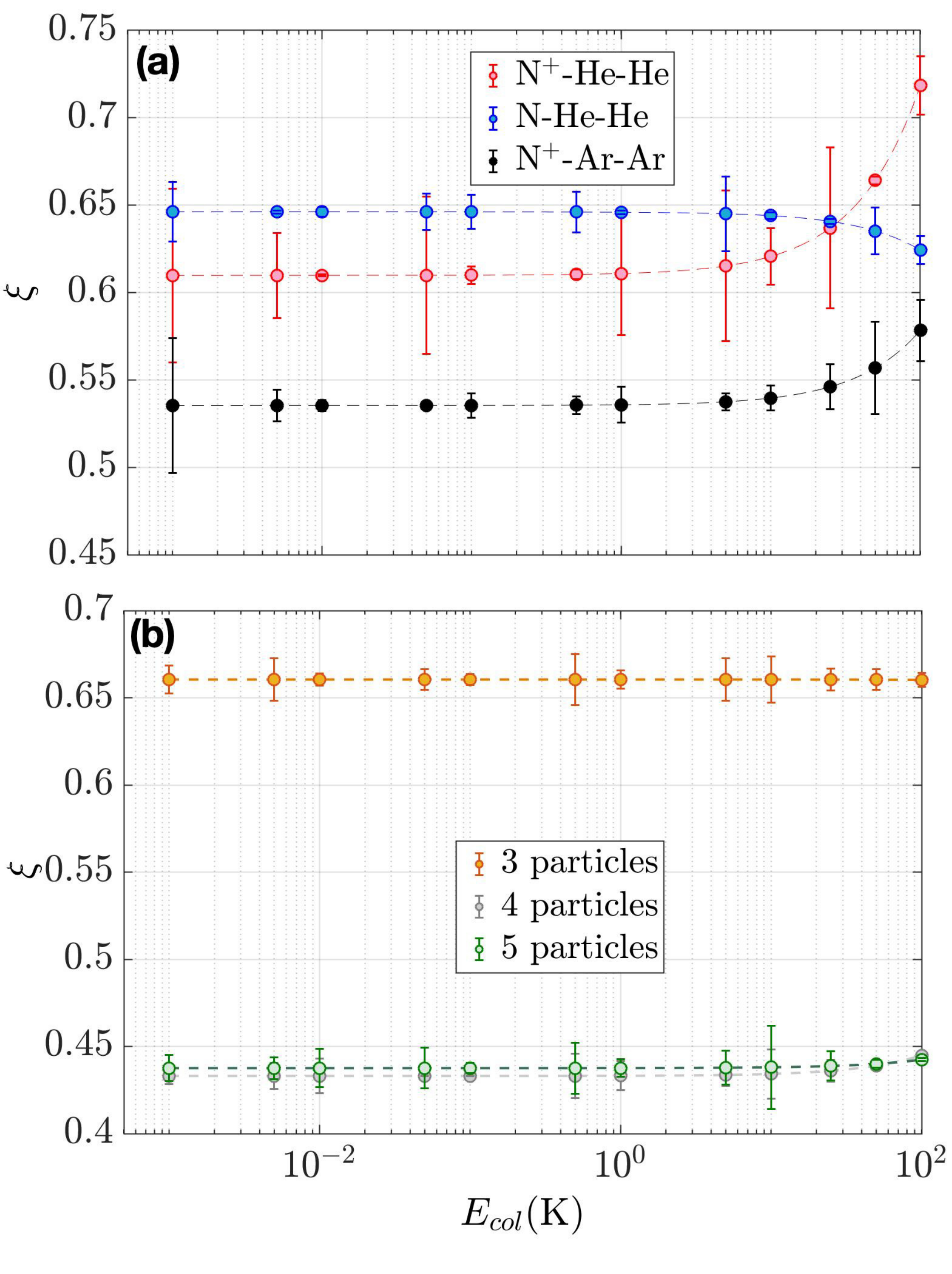}
		\caption{Shape parameter $\xi$ of the hyperradial distribution as a function of the collision energy. Panel (a) shows the same few-body systems as in \cref{fig2}, whereas panel (b) is associated with the systems presented in \cref{fig3}. The error bars represent the fitting uncertainty on this parameter. A dashed line accompanies the data points to guide the eye.}
		\label{fig5}
	\end{figure}

	Next, we analyze the parameters of the GEV as a function of the collision energy for a different number of particles and interparticle interactions. In particular, the collision energies spanned between 1mK and 100K, thus covering collisions between the cold and the thermal regimes. \cref{fig5} displays the shape parameter, $\xi$, as a function of the collision energy for  different few-body systems. In particular, panel (a) shows the three different three-body systems as in \cref{fig2}, whereas panel (b) covers the systems displayed in \cref{fig3}, i.e., N-body systems with $N=3,4,5$. It is worth noting that, independently of the interparticle interaction and number of particles,  $\xi$ is positive, thus, the GEV distribution is a Fréchet type ($\xi>0$)  in all cases. Furthermore, $\xi$ is energy independent, i.e., the shape of the hyperradial distribution remains the same for all collision energies within a given interaction model, i.e., with the same N and same interparticle interactions. In addition, we notice two different trends in panel~(a): while at the energies below 1~K the shape parameter is constant, for $E_\text{col}\gtrsim$ 1~K it increases or decreases monotonically. 
	
\begin{figure}[h]
		\centering
		\includegraphics[scale=0.4]{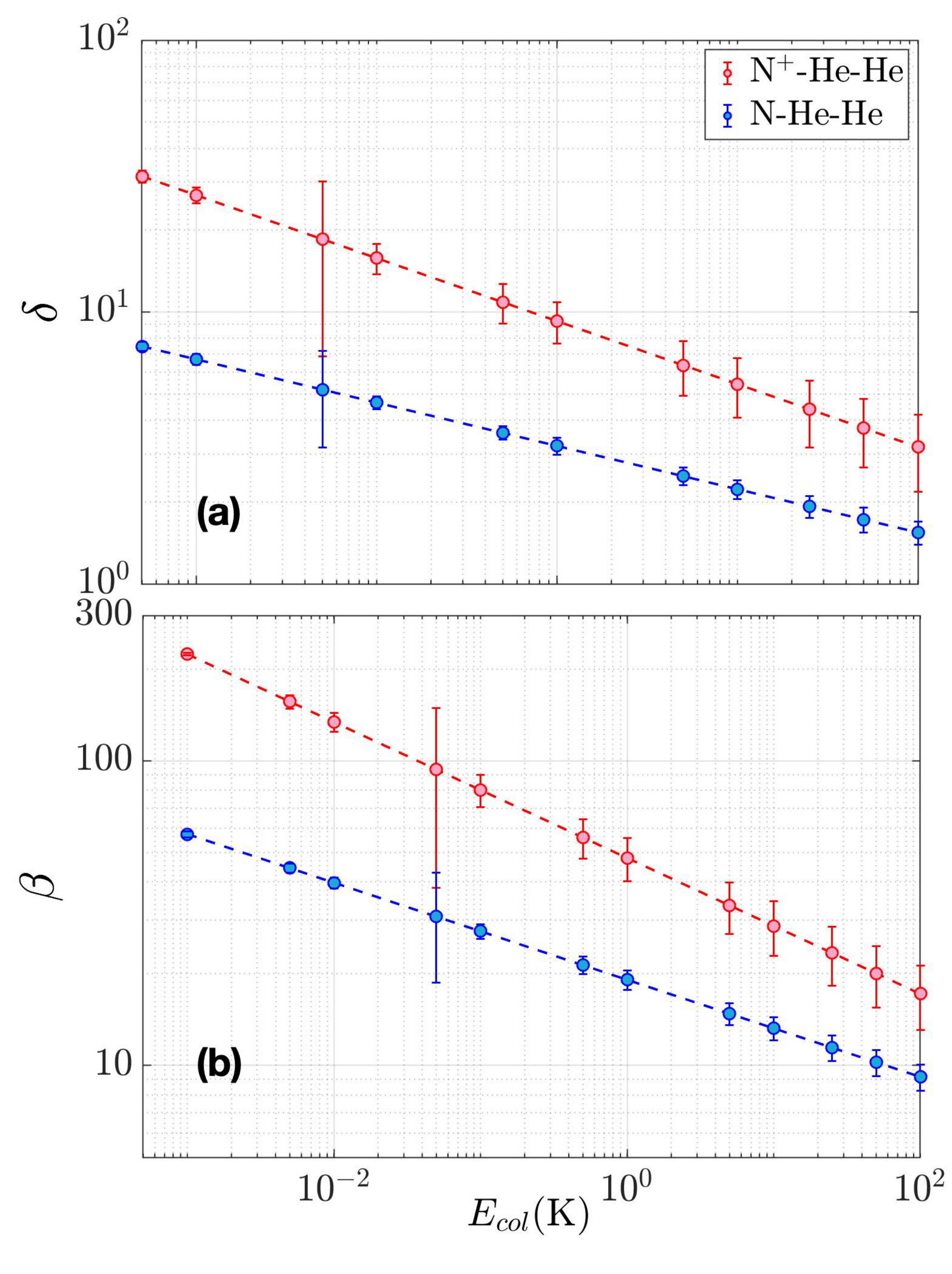}
		\caption{$\delta$ [panel (a)] and $\beta$  [panel (b)] hyperradial distribution parameters as a function of the collision energy. The error bars represent the fitting uncertainty on this parameter. The dashed-lines is the best fit of the data points to a power-law function, $a E_{\text{col}}^b$.}
		\label{fig6}
	\end{figure}

	The study on the relationship between $\delta$ and $\beta$ parameters of the GEV distribution with the collision energy is shown in \cref{fig6}, for three-body systems presenting two different types of interparticle interactions. In this figure, it is noticed that, as opposed to the results for the parameter $\xi$, these parameters have a strong dependence on the collision energy. In particular, these parameters have a power-law relationship with the collision energy. To study this in detail, we have extended our analysis to different few-body systems with different interactions, and the corresponding parameters of the GEV distribution ($\xi,\beta$ and $\delta$) are listed in \cref{table1}. In accordance with the results depicted in \cref{fig5}, we notice that for each system under consideration parameter $\xi$ is almost independent of the collision energy (the maximum and minimum values of the variation interval are listed in the second and third columns of \cref{table1}). On the other hand, we find that $\beta$ and $\delta$ follow a power-law relationship with the collision energy as $a\times E_{\text{col}}^b$, independently of the interparticle interaction. Intriguingly, we only observe two powers: $b=- \frac{1}{6}$ for vdW-type interactions and $b=-\frac{1}{4}$ for systems containing charged-neutral interactions, independently of the number of particles. Therefore, assuming that $\xi$ is energy independent, the mode of the distribution given by 
	\begin{equation}
		\mathbf{\rho}_m=\beta + \frac{\delta}{\xi}((1+\xi)^{-\xi}-1),   
	\end{equation}
	yields $\rho_m\propto E_{\text{col}}^{-\frac{1}{6}}$ and $\rho_m\propto E_{\text{col}}^{-\frac{1}{4}}$ for systems including vdW-type interactions and charged-neutral interactions, respectively. In other words, the most probable value of the capture hyperradius only depends on the long-range tail of the dominant inter-particle interactions, as it has been argued before~\cite{Perez-Rios2014,Perez-Rios2015,Perez-Rios2018}, but never specifically proved.

	\begin{table}[t]
		\begin{center}
			\begin{minipage}{\textwidth}
				\caption{Relationship between parameters of the hyperradial distribution function and the collision energy for few-body systems including vdW-type and charged-neutral interactions. Note: $\xi$ only varies from system to system but it is independent of the collision energy, and  $\beta$ and $\delta$ follow a power-law relationship as a function of the collision energy as $a \times E_{\text{col}}^b$}
				\begin{tabular*}{\textwidth}{@{\extracolsep{\fill}}lcccccc@{\extracolsep{\fill}}}
					\toprule%
					& \multicolumn{2}{@{}c@{}}{$\xi$} & \multicolumn{2}{@{}c@{}}{$\beta$} & \multicolumn{2}{@{}c@{}}{$\delta$}\\\cmidrule{2-3}\cmidrule{4-5}\cmidrule{6-7}%
					System & Min & Max& a & b & a & b \\
					\midrule
					N-He-He  & 0.610 & 0.667 & 9.110$\pm$3.165 & -0.159$\pm$0.032 & 3.216$\pm$0.525 & -0.159$\pm$0.031\\
					N$^+$-He-He  & 0.560 & 0.700 & 47.99$\pm$14.84 & -0.223$\pm$0.056 & 6.941$\pm$0.136 & -0.232$\pm$0.060\\
					N$^+$-3He  & 0.319 & 0.403 & 63.67$\pm$19.83 & -0.222$\pm$0.055 & 19.06$\pm$6.435 & -0.229$\pm$0.06\\
					N$^+$-4He  & 0.327 & 0.422 & 75.37$\pm$22.23 & -0.272$\pm$0.053 & 20.61$\pm$6.62 & -0.229$\pm$0.060\\
					N$^+$-Ar-Ar  & 0.487 & 0.583 & 102.4$\pm$31.6 & -0.224$\pm$0.056 & 21.35$\pm$6.85 & -0.231$\pm$0.057\\
					As-He-He  & 0.674 & 0.646 & 21.79$\pm$0.09 & -0.166 & 3.764$\pm$0.044 & -0.166$\pm$0.002\\
					As-3He  & 0.423 & 0.446 & 36.6$\pm$6.07 & -0.160 $\pm$ 0.035 & 9.471$\pm$1.578 & -0.159$\pm$0.033\\
					As-4He  & 0.416 & 0.446 & 45.9$\pm$7.49 & -0.159$\pm$ 0.319 & 11.36$\pm$1.85 & -0.158$\pm$0.032\\
					\botrule
				\end{tabular*}
				\label{table1}
			\end{minipage}
		\end{center}
	\end{table} 
		
	\subsection{Fréchet Distribution}
	All the simulations above show that capture hyperradial distribution follows the Fréchet distribution within the GEV family, i.e., $\xi>0$. Throughout this section, we try to obtain these results independent of our observation and using the extreme value theory.
	
	Each point in the D-dimensional space, indicated by a hyperradius and D-1 hyperangles, corresponds to a set of all possible configurations of the N particles in the 3-dimensional space. The hyperangles specify the shape of the configuration (e.g., in the 3-body case, it is a triangle). In contrast, the hyperradius describes the size (in a geometrical way) of the N-body system (the enlarging or shrinking of the atomic configuration by the same scale factor in all directions).
	
	As explained above, to find the capture hyperradius, a set of hyperangles ($\vec{\alpha}$) are randomly generated; those define the shape of the atomic configuration. Once the hyperangles are chosen, the hyperradius, as free of constraints, can be chosen randomly from a set of independent and identically distributed variables. In particular, by generating each set of $\vec{\alpha}$ a sequence of hyperradius are available, and by solving \cref{eq8}, we choose the largest $\rho$. Finally, repeating this procedure for a large enough number of Monte Carlo steps, we generate a distribution of capture hyperradius.  
	
	According to the Fisher-Tippett-Gnedenko theorem~\cite{10.2307/1968974,fisher_tippett_1928,Leadbetter2012,LaurensdeHaan2007}, the limiting distribution of such variables (the extreme value distribution) after proper normalization is either Gumbel,  Fréchet, or Weibull distribution. These three distributions have different behavior over their range~\cite{Singh1998,Leadbetter2012}. The convergence towards the Weibull distribution requires the existence of an upper bound (finite number) on the support of the distribution. However, it is always possible to find configurations satisfying \cref{eq8} in which $\rho$ is very large. Therefore, the distribution can not be Weibull. The Gumbel distribution does not have finite upper or lower bounds, implying, in our case, that the size of the $N$ particle configuration can be arbitrarily small and still satisfying \cref{eq8}, which is not possible. The range of the Fréchet distribution is bounded from below; thus, this is the only distribution that can accurately describe the capture hyperradius distribution.  
	
	\section{Conclusion}\label{sec5}
	The study of classical capture models relies on defining the capture radius, which in the case of few-body systems translates into the distribution of hyperradius when the system is described in hyperspherical coordinates. This work has computed the hyperradial distribution for three- four- and five-body systems, including vdW and charged-neutral interactions in a wide range of collision energies between the cold and the thermal regimes. As a result, we conclude that the hyperradial distribution function follows a Fréchet distribution independently of the interparticle interaction nature and the number of interacting particles. In other words, classical capture models will lead to a Fréchet distribution in any few-body system.
	
	Finally, after using extreme value theory, we have found a plausible explanation behind the universality of the Fréchet distribution for the capture hyperradius, thus finding an application of extreme value theory in few-body physics.

	\begin{acknowledgments}
	
	M. M acknowledges the support of the Deutsche Forschungsgemeinschaft (DFG – German Research Foundation) under the grant number PE 3477/2 - 493725479.  J. P.-R. acknowledges the support of the Simons Foundations and the Deutsche Forschungsgemeinschaft (DFG – German Research Foundation) under the grant number PE 3477/2 - 493725479. A A. E acknowledges the support of the AFOSR-MURI under grant
	number FA9550-20-1-0323. 
   \end{acknowledgments}

	\bibliographystyle{apsrev4-2}
\bibliography{HRD.bib}

\begin{thebibliography}{25}%
\makeatletter
\providecommand \@ifxundefined [1]{%
 \@ifx{#1\undefined}
}%
\providecommand \@ifnum [1]{%
 \ifnum #1\expandafter \@firstoftwo
 \else \expandafter \@secondoftwo
 \fi
}%
\providecommand \@ifx [1]{%
 \ifx #1\expandafter \@firstoftwo
 \else \expandafter \@secondoftwo
 \fi
}%
\providecommand \natexlab [1]{#1}%
\providecommand \enquote  [1]{``#1''}%
\providecommand \bibnamefont  [1]{#1}%
\providecommand \bibfnamefont [1]{#1}%
\providecommand \citenamefont [1]{#1}%
\providecommand \href@noop [0]{\@secondoftwo}%
\providecommand \href [0]{\begingroup \@sanitize@url \@href}%
\providecommand \@href[1]{\@@startlink{#1}\@@href}%
\providecommand \@@href[1]{\endgroup#1\@@endlink}%
\providecommand \@sanitize@url [0]{\catcode `\\12\catcode `\$12\catcode
  `\&12\catcode `\#12\catcode `\^12\catcode `\_12\catcode `\%12\relax}%
\providecommand \@@startlink[1]{}%
\providecommand \@@endlink[0]{}%
\providecommand \url  [0]{\begingroup\@sanitize@url \@url }%
\providecommand \@url [1]{\endgroup\@href {#1}{\urlprefix }}%
\providecommand \urlprefix  [0]{URL }%
\providecommand \Eprint [0]{\href }%
\providecommand \doibase [0]{https://doi.org/}%
\providecommand \selectlanguage [0]{\@gobble}%
\providecommand \bibinfo  [0]{\@secondoftwo}%
\providecommand \bibfield  [0]{\@secondoftwo}%
\providecommand \translation [1]{[#1]}%
\providecommand \BibitemOpen [0]{}%
\providecommand \bibitemStop [0]{}%
\providecommand \bibitemNoStop [0]{.\EOS\space}%
\providecommand \EOS [0]{\spacefactor3000\relax}%
\providecommand \BibitemShut  [1]{\csname bibitem#1\endcsname}%
\let\auto@bib@innerbib\@empty
\bibitem [{\citenamefont {Levine}(2005)}]{Levine2005}%
  \BibitemOpen
  \bibfield  {author} {\bibinfo {author} {\bibfnamefont {R.}~\bibnamefont
  {Levine}},\ }\href@noop {} {\emph {\bibinfo {title} {Molecular reaction
  dynamics}}}\ (\bibinfo  {publisher} {Cambridge University Press},\ \bibinfo
  {address} {Cambridge, UK New York},\ \bibinfo {year} {2005})\BibitemShut
  {NoStop}%
\bibitem [{\citenamefont {P{\'{e}}rez-R{\'{i}}os}(2020)}]{Perez-Rios2020}%
  \BibitemOpen
  \bibfield  {author} {\bibinfo {author} {\bibfnamefont {J.}~\bibnamefont
  {P{\'{e}}rez-R{\'{i}}os}},\ }\href@noop {} {\emph {\bibinfo {title} {An
  Introduction to Cold and Ultracold Chemistry}}}\ (\bibinfo  {publisher}
  {Springer International Publishing},\ \bibinfo {address} {Cham,
  Switzerland},\ \bibinfo {year} {2020})\BibitemShut {NoStop}%
\bibitem [{Note1()}]{Note1}%
  \BibitemOpen
  \bibinfo {note} {This is strictly true in the case of zero impact parameter,
  or null angular momentum collision.}\BibitemShut {Stop}%
\bibitem [{\citenamefont {Langevin}(1905)}]{Langevin1905}%
  \BibitemOpen
  \bibfield  {author} {\bibinfo {author} {\bibfnamefont {M.~P.}\ \bibnamefont
  {Langevin}},\ }\href
  {https://gallica.bnf.fr/ark:/12148/bpt6k34935p/f242.item.texteImage?lang=EN}
  {\bibfield  {journal} {\bibinfo  {journal} {Ann. Chim. Phys}\ }\textbf
  {\bibinfo {volume} {5}},\ \bibinfo {pages} {245} (\bibinfo {year}
  {1905})}\BibitemShut {NoStop}%
\bibitem [{\citenamefont {P{\'{e}}rez-R{\'{i}}os}\ \emph
  {et~al.}(2014)\citenamefont {P{\'{e}}rez-R{\'{i}}os}, \citenamefont {Ragole},
  \citenamefont {Wang},\ and\ \citenamefont {Greene}}]{Perez-Rios2014}%
  \BibitemOpen
  \bibfield  {author} {\bibinfo {author} {\bibfnamefont {J.}~\bibnamefont
  {P{\'{e}}rez-R{\'{i}}os}}, \bibinfo {author} {\bibfnamefont {S.}~\bibnamefont
  {Ragole}}, \bibinfo {author} {\bibfnamefont {J.}~\bibnamefont {Wang}},\ and\
  \bibinfo {author} {\bibfnamefont {C.~H.}\ \bibnamefont {Greene}},\ }\href
  {https://doi.org/10.1063/1.4861851} {\bibfield  {journal} {\bibinfo
  {journal} {J. Chem. Phys.}\ }\textbf {\bibinfo {volume} {140}},\ \bibinfo
  {pages} {044307} (\bibinfo {year} {2014})}\BibitemShut {NoStop}%
\bibitem [{\citenamefont {P{\'{e}}rez-R{\'{i}}os}\ and\ \citenamefont
  {Greene}(2018)}]{Perez-Rios2018}%
  \BibitemOpen
  \bibfield  {author} {\bibinfo {author} {\bibfnamefont {J.}~\bibnamefont
  {P{\'{e}}rez-R{\'{i}}os}}\ and\ \bibinfo {author} {\bibfnamefont {C.~H.}\
  \bibnamefont {Greene}},\ }\href {https://doi.org/10.1103/PhysRevA.98.062707}
  {\bibfield  {journal} {\bibinfo  {journal} {Phys. Rev. A}\ }\textbf {\bibinfo
  {volume} {98}},\ \bibinfo {pages} {23} (\bibinfo {year} {2018})}\BibitemShut
  {NoStop}%
\bibitem [{\citenamefont {P{\'{e}}rez-R{\'{i}}os}(2021)}]{Perez-Rios2021}%
  \BibitemOpen
  \bibfield  {author} {\bibinfo {author} {\bibfnamefont {J.}~\bibnamefont
  {P{\'{e}}rez-R{\'{i}}os}},\ }\href
  {https://doi.org/10.1080/00268976.2021.1881637} {\bibfield  {journal}
  {\bibinfo  {journal} {Mol. Phys.}\ }\textbf {\bibinfo {volume} {119}},\
  \bibinfo {pages} {e1881637} (\bibinfo {year} {2021})}\BibitemShut {NoStop}%
\bibitem [{\citenamefont {Mirahmadi}\ and\ \citenamefont
  {P{\'{e}}rez-R{\'{i}}os}(2021)}]{Mirahmadi2021}%
  \BibitemOpen
  \bibfield  {author} {\bibinfo {author} {\bibfnamefont {M.}~\bibnamefont
  {Mirahmadi}}\ and\ \bibinfo {author} {\bibfnamefont {J.}~\bibnamefont
  {P{\'{e}}rez-R{\'{i}}os}},\ }\href {https://doi.org/10.1063/5.0039610}
  {\bibfield  {journal} {\bibinfo  {journal} {J. Chem. Phys.}\ }\textbf
  {\bibinfo {volume} {154}},\ \bibinfo {pages} {034305} (\bibinfo {year}
  {2021})}\BibitemShut {NoStop}%
\bibitem [{\citenamefont {Mirahmadi}\ and\ \citenamefont
  {P{\'e}rez-R{\'\i}os}(2021)}]{Mirahmadi2021b}%
  \BibitemOpen
  \bibfield  {author} {\bibinfo {author} {\bibfnamefont {M.}~\bibnamefont
  {Mirahmadi}}\ and\ \bibinfo {author} {\bibfnamefont {J.}~\bibnamefont
  {P{\'e}rez-R{\'\i}os}},\ }\href {https://doi.org/10.1063/5.0062812}
  {\bibfield  {journal} {\bibinfo  {journal} {The Journal of Chemical Physics}\
  }\textbf {\bibinfo {volume} {155}},\ \bibinfo {pages} {094306} (\bibinfo
  {year} {2021})}\BibitemShut {NoStop}%
\bibitem [{\citenamefont {Singh}(1998)}]{Singh1998}%
  \BibitemOpen
  \bibfield  {author} {\bibinfo {author} {\bibfnamefont {V.~P.}\ \bibnamefont
  {Singh}},\ }\href@noop {} {\emph {\bibinfo {title} {Entropy-Based Parameter
  Estimation in Hydrology}}}\ (\bibinfo  {publisher} {Springer Netherlands},\
  \bibinfo {address} {Dordrecht},\ \bibinfo {year} {1998})\BibitemShut
  {NoStop}%
\bibitem [{\citenamefont {Smith}(1960)}]{Smith1960}%
  \BibitemOpen
  \bibfield  {author} {\bibinfo {author} {\bibfnamefont {F.~T.}\ \bibnamefont
  {Smith}},\ }\href {https://doi.org/10.1103/PhysRev.120.1058} {\bibfield
  {journal} {\bibinfo  {journal} {Phys. Rev.}\ }\textbf {\bibinfo {volume}
  {120}},\ \bibinfo {pages} {1058} (\bibinfo {year} {1960})}\BibitemShut
  {NoStop}%
\bibitem [{\citenamefont {Smith}(1962)}]{Smith1962}%
  \BibitemOpen
  \bibfield  {author} {\bibinfo {author} {\bibfnamefont {F.~T.}\ \bibnamefont
  {Smith}},\ }\href {https://doi.org/10.1039/DF9623300183} {\bibfield
  {journal} {\bibinfo  {journal} {Discuss. Faraday Soc.}\ }\textbf {\bibinfo
  {volume} {33}},\ \bibinfo {pages} {183} (\bibinfo {year} {1962})}\BibitemShut
  {NoStop}%
\bibitem [{\citenamefont {Shui}(1972)}]{Shui1972}%
  \BibitemOpen
  \bibfield  {author} {\bibinfo {author} {\bibfnamefont {V.~H.}\ \bibnamefont
  {Shui}},\ }\href {https://doi.org/10.1063/1.1678459} {\bibfield  {journal}
  {\bibinfo  {journal} {The Journal of Chemical Physics}\ }\textbf {\bibinfo
  {volume} {57}},\ \bibinfo {pages} {1704} (\bibinfo {year}
  {1972})}\BibitemShut {NoStop}%
\bibitem [{\citenamefont {Shui}(1973)}]{Shui1973}%
  \BibitemOpen
  \bibfield  {author} {\bibinfo {author} {\bibfnamefont {V.~H.}\ \bibnamefont
  {Shui}},\ }\href {https://doi.org/10.1063/1.1679071} {\bibfield  {journal}
  {\bibinfo  {journal} {The Journal of Chemical Physics}\ }\textbf {\bibinfo
  {volume} {58}},\ \bibinfo {pages} {4868} (\bibinfo {year}
  {1973})}\BibitemShut {NoStop}%
\bibitem [{\citenamefont {Avery}(2012)}]{Avery2012}%
  \BibitemOpen
  \bibfield  {author} {\bibinfo {author} {\bibfnamefont {J.}~\bibnamefont
  {Avery}},\ }\href@noop {} {\emph {\bibinfo {title} {Hyperspherical Harmonics:
  Applications in Quantum Theory}}},\ Reidel Texts in the Mathematical
  Sciences\ (\bibinfo  {publisher} {Springer Netherlands},\ \bibinfo {address}
  {Dordrecht},\ \bibinfo {year} {2012})\BibitemShut {NoStop}%
\bibitem [{\citenamefont {Landau}\ and\ \citenamefont
  {Binder}(2009)}]{landau_binder_2009}%
  \BibitemOpen
  \bibfield  {author} {\bibinfo {author} {\bibfnamefont {D.~P.}\ \bibnamefont
  {Landau}}\ and\ \bibinfo {author} {\bibfnamefont {K.}~\bibnamefont
  {Binder}},\ }\href {https://doi.org/10.1017/CBO9780511994944} {\emph
  {\bibinfo {title} {A Guide to Monte Carlo Simulations in Statistical
  Physics}}},\ \bibinfo {edition} {3rd}\ ed.\ (\bibinfo  {publisher} {Cambridge
  University Press},\ \bibinfo {year} {2009})\BibitemShut {NoStop}%
\bibitem [{\citenamefont {Shreider}(2014)}]{MC2}%
  \BibitemOpen
  \bibfield  {author} {\bibinfo {author} {\bibfnamefont {Y.~A.}\ \bibnamefont
  {Shreider}},\ }\href@noop {} {\emph {\bibinfo {title} {The Monte Carlo
  method: the method of statistical trials}}},\ Vol.~\bibinfo {volume} {87}\
  (\bibinfo  {publisher} {Elsevier},\ \bibinfo {address} {Oxford},\ \bibinfo
  {year} {2014})\BibitemShut {NoStop}%
\bibitem [{\citenamefont {Aziz}\ \emph {et~al.}(1995)\citenamefont {Aziz},
  \citenamefont {Janzen},\ and\ \citenamefont {Moldover}}]{C6He2}%
  \BibitemOpen
  \bibfield  {author} {\bibinfo {author} {\bibfnamefont {R.~A.}\ \bibnamefont
  {Aziz}}, \bibinfo {author} {\bibfnamefont {A.~R.}\ \bibnamefont {Janzen}},\
  and\ \bibinfo {author} {\bibfnamefont {M.~R.}\ \bibnamefont {Moldover}},\
  }\href {https://doi.org/10.1103/PhysRevLett.74.1586} {\bibfield  {journal}
  {\bibinfo  {journal} {Phys. Rev. Lett.}\ }\textbf {\bibinfo {volume} {74}},\
  \bibinfo {pages} {1586} (\bibinfo {year} {1995})}\BibitemShut {NoStop}%
\bibitem [{\citenamefont {Kumar}\ and\ \citenamefont {Meath}(1985)}]{C6Ar2}%
  \BibitemOpen
  \bibfield  {author} {\bibinfo {author} {\bibfnamefont {A.}~\bibnamefont
  {Kumar}}\ and\ \bibinfo {author} {\bibfnamefont {W.~J.}\ \bibnamefont
  {Meath}},\ }\href {https://doi.org/10.1080/00268978500103191} {\bibfield
  {journal} {\bibinfo  {journal} {Molecular Physics}\ }\textbf {\bibinfo
  {volume} {54}},\ \bibinfo {pages} {823} (\bibinfo {year} {1985})}\BibitemShut
  {NoStop}%
\bibitem [{\citenamefont {Partridge}\ \emph {et~al.}(2001)\citenamefont
  {Partridge}, \citenamefont {Stallcop},\ and\ \citenamefont {Levin}}]{C6NHe}%
  \BibitemOpen
  \bibfield  {author} {\bibinfo {author} {\bibfnamefont {H.}~\bibnamefont
  {Partridge}}, \bibinfo {author} {\bibfnamefont {J.~R.}\ \bibnamefont
  {Stallcop}},\ and\ \bibinfo {author} {\bibfnamefont {E.}~\bibnamefont
  {Levin}},\ }\href {https://doi.org/10.1063/1.1385372} {\bibfield  {journal}
  {\bibinfo  {journal} {The Journal of Chemical Physics}\ }\textbf {\bibinfo
  {volume} {115}},\ \bibinfo {pages} {6471} (\bibinfo {year}
  {2001})}\BibitemShut {NoStop}%
\bibitem [{\citenamefont {P{\'{e}}rez-R{\'{i}}os}\ and\ \citenamefont
  {Greene}(2015)}]{Perez-Rios2015}%
  \BibitemOpen
  \bibfield  {author} {\bibinfo {author} {\bibfnamefont {J.}~\bibnamefont
  {P{\'{e}}rez-R{\'{i}}os}}\ and\ \bibinfo {author} {\bibfnamefont {C.~H.}\
  \bibnamefont {Greene}},\ }\href {https://doi.org/10.1063/1.4927702}
  {\bibfield  {journal} {\bibinfo  {journal} {J. Chem. Phys.}\ }\textbf
  {\bibinfo {volume} {143}},\ \bibinfo {pages} {041105} (\bibinfo {year}
  {2015})}\BibitemShut {NoStop}%
\bibitem [{\citenamefont {Gnedenko}(1943)}]{10.2307/1968974}%
  \BibitemOpen
  \bibfield  {author} {\bibinfo {author} {\bibfnamefont {B.}~\bibnamefont
  {Gnedenko}},\ }\href {http://www.jstor.org/stable/1968974} {\bibfield
  {journal} {\bibinfo  {journal} {Annals of Mathematics}\ }\textbf {\bibinfo
  {volume} {44}},\ \bibinfo {pages} {423} (\bibinfo {year} {1943})}\BibitemShut
  {NoStop}%
\bibitem [{\citenamefont {Fisher}\ and\ \citenamefont
  {Tippett}(1928)}]{fisher_tippett_1928}%
  \BibitemOpen
  \bibfield  {author} {\bibinfo {author} {\bibfnamefont {R.~A.}\ \bibnamefont
  {Fisher}}\ and\ \bibinfo {author} {\bibfnamefont {L.~H.~C.}\ \bibnamefont
  {Tippett}},\ }\href {https://doi.org/10.1017/S0305004100015681} {\bibfield
  {journal} {\bibinfo  {journal} {Mathematical Proceedings of the Cambridge
  Philosophical Society}\ }\textbf {\bibinfo {volume} {24}},\ \bibinfo {pages}
  {180–190} (\bibinfo {year} {1928})}\BibitemShut {NoStop}%
\bibitem [{\citenamefont {Leadbetter}\ \emph {et~al.}(2012)\citenamefont
  {Leadbetter}, \citenamefont {Lindgren},\ and\ \citenamefont
  {Rootzen}}]{Leadbetter2012}%
  \BibitemOpen
  \bibfield  {author} {\bibinfo {author} {\bibfnamefont {M.~R.}\ \bibnamefont
  {Leadbetter}}, \bibinfo {author} {\bibfnamefont {G.}~\bibnamefont
  {Lindgren}},\ and\ \bibinfo {author} {\bibfnamefont {H.}~\bibnamefont
  {Rootzen}},\ }\href@noop {} {\emph {\bibinfo {title} {Extremes and Related
  Properties of Random Sequences and Processes}}}\ (\bibinfo  {publisher}
  {Springer},\ \bibinfo {address} {New York},\ \bibinfo {year}
  {2012})\BibitemShut {NoStop}%
\bibitem [{\citenamefont {Laurens~de Haan}(2007)}]{LaurensdeHaan2007}%
  \BibitemOpen
  \bibfield  {author} {\bibinfo {author} {\bibfnamefont {A.~F.}\ \bibnamefont
  {Laurens~de Haan}},\ }\href@noop {} {\emph {\bibinfo {title} {Extreme Value
  Theory}}}\ (\bibinfo  {publisher} {Springer-Verlag GmbH},\ \bibinfo {address}
  {New York},\ \bibinfo {year} {2007})\BibitemShut {NoStop}%
\end{thebibliography}%
\end{document}